\RequirePackage{fix-cm}
\documentclass[5p,times]{elsarticle}
\usepackage{graphicx}
\usepackage{subfigure}
\usepackage{epsfig}
\usepackage{color}
\usepackage{url}
\usepackage{comment}
\usepackage{pstricks}
\usepackage{pict2e,color,graphics}
\usepackage{natbib}
\usepackage{appendix}

\begin{document}

\hyphenation{cen-tral-i-sa-tion}
\hyphenation{cen-tralised}

\begin{frontmatter}
\title{Organisations (de-)centralised to a greater or lesser degree for allocating cities in two Multiple Travelling Salesmen Problems}


\author{Thierry Moyaux}
\address{Univ Lyon, INSA Lyon, Univ Jean Monnet Saint-Etienne, Université Claude Bernard Lyon 1, Univ Lyon 2, DISP-UR 4570, Villeurbanne G-69621, France\\
              Tel.: (+33) 4 72 43 75 37}
\ead{Thierry.Moyaux@insa-lyon.fr}




\begin{abstract}
Decisions in organisations may be made either by a Central Authority (CA), $e.g.,$ in a hierarchy, or by the agents in a decentralised way, $e.g.,$ in a heterarchy. Since both kinds of organisations have their advantages ($e.g.,$ optimality for centralised organisations and reactivity for decentralised ones), our goal is ultimately to understand when and how to use each of them.
Our previous work proposed a variant of the Multiple Travelling Salesmen Problem, which we now call MTSP$_\textrm{s}$. We use the subscript ``s'' to refer to salesmen's selfishness when they minimise their individual route length. If, on the contrary, they are assumed to be benevolent, we add subscript ``b''and thus the term MTSP$_\textrm{b}$ to refer to the traditional MTSP in which the salesmen minimise the total route length.
This article shows how to obtain such benevolent agents by slightly modifying selfish agents.
We can then compare organisations which are (de-)centralised to a greater or lesser degree, which enables us to carry out the allocation of cities in the MTSP$_\textrm{b}$.
The first experiment shows that the relative efficiency (ranking) of the organisations differs between MTSP$_\textrm{b}$ and MTSP$_\textrm{s}$.
Since reactivity fosters decentralisation, the second experiment gradually reduces the time taken for it to impact this ranking. Both experiments show that pure centralisation is either the best or the worst option, and that the zone between the two situations is very narrow.

\end{abstract}

\begin{keyword}
Decision making \sep Centralised decision \sep Decentralised decision \sep MTSP
\end{keyword}


\end{frontmatter}

\section{Introduction}
\label{sec:introduction}
The structure of an organisation may be (de-)centralised to a greater or lesser degree. We call ``organisation'' the structure of this organisation, and ``mechanism'' an instance of organisation. Our goal is ultimately to establish which structure is best suited to a situation.
To this end, this article compares organisations that are (de-)centralised to a greater or lesser degree, to solve the MTSP (Multiple Travelling Salesmen Problem), $i.e.$, the Vehicle Routing Problem without capacity constraints \citep{toth02, bektas06}. This problem consists of two connected subproblems: allocation and routing. This article studies the (de-)centralisation of the former.
In this context, our research question may be stated as: when is it more efficient to let salespersons/agents concurrently solve many small Travelling Salesman Problems (TSP, $i.e.,$ the routing subproblem alone), rather than letting a Central Authority (CA) solve the single large Multiple Travelling Salesmen Problem (MTSP, $i.e.,$ both allocation and routing problems together)?
We call the centralised organisation {\sf Centr}. The others are decentralised organisations and we will see that they have various levels of decentralisation.

This research question is important because optimising the operations in a bad organisation may not be as efficient as improving the organisation itself. For example, (de-)centralising the maintenance of buildings in a university seems to be a completely different question from that of (de-)centralising the maintenance of its computers, as the required reactivity of these two types of maintenance is very different. Repairing lights, doors, toilets, roofs, and so forth can wait for a few hours without completely bringing to a halt teaching, research and administration, whereas a failure of computers hardware or software disturbs such operations significantly.
When an organisation is in place, it is very difficult to change it because this would impact:
(i) technologies supporting decisions, $i.e.,$ the information system in terms of structure, technologies ($e.g.,$ need for big servers and/or smaller devices such as laptops and smartphones), tools, etc.;
(ii) resistance to change (power -- the Central Authority (CA) in a centralised organisation will never give up their power to switch to a decentralised organisation -- and habits);
(iii) loss of experience (processes and tools previously optimised will be replaced by new unoptimised ones);
(iv) etcetera.

This question has been addressed by various communities, as our literature review in Economics and Computer Science literature review attests \citep{moyaux12}. In particular, this review points to a comparison of four organisations to solve a resource allocation problem for production and transportation by \cite{davidsson07} which, to our knowledge, is the only quantified comparison of as many as four organisations.
More recently, \cite{cardin17} have reviewed the literature on Holonic Control Architectures and concluded that additional research on the coupling of centralised and decentralised mechanisms in a same organisation should address the following three problems:
(i) the estimation of future performance because a disturbance may have either no impact or a huge impact on the system;
(ii) the design of switching indicators which use the estimation in i to decide if and when to switching from a centralised to a decentralised mechanism; and
(iii) the strategy for switching between both mechanisms seamlessly without stopping the system.
Our work addresses the first problem.
It moreover resembles that of \cite{zhang19}, but in the area of transportation rather than production scheduling. More precisely, these authors review traditional (centralised) scheduling in order to propose future research direction for ``smart distributed scheduling'' in Industry 4.0.
In fact, \cite{derigent20} identify ``Autonomous and decentralised decision support systems'' as one of the ten key enablers of Industry 4.0.
As noted above, some of our organisations are not purely (de-)centralised, which is also investigated in the area of manufacturing systems by \cite{sallez10}, among others.
Concerning MTSP, \cite{kivelevitch13} propose a very interesting decentralised mechanism -- different from ours -- for the organisation which we call {\sf P2P$_b$}, that is, Peer to Peer with benevolent agents.

In this article we propose to take a centralised problem that has been thoroughly investigated by operational research and to study when and how a combination of operational research and multiagent techniques may decentralise it efficiently. Since operational research is able to guarantee the optimality of a solution when enough time is available, we set a time limit in order to quantify the reactivity of organisations. This is consistent with the intuition that decentralisation is more efficient when reactivity constraints the time available to make decisions.
From a more technical perspective, we explore how agents/salespersons may solve the MTSP in organisations with a greater or lesser degree of (de-)centralisation. We call the traditional version of this problem MTSP$_\textrm{b}$, for MTSP with benevolent agents, in the sense that they minimise the sum of the individual route lengths of all the agents, that is, the total route length. In contrast, our previous article \citep{moyaux20} proposes MTSP$_\textrm{s}$ in which the agents are selfish as they want to minimise their individual route length only.
Our \emph{contributions} may be summarised as follows:
\begin{enumerate}
	\item \emph{Quantification of the efficiency of organisations that are (de-)centralised to a greater or lesser degree}: Our experimental results show the performance of each mechanism when it solves a set of instances compared to the performance of the centralised organisation when it solves the same instances. More precisely, we show the median and ninth decile of the ratios of the total route length found by a mechanism on these instances divided by the total route length found by ``{\sf Centr$_b$}''. As noted above, we set a limit to the computation time such that this ratio may be lower than 1 when {\sf Centr$_b$} does not have enough time to find the best allocation and routings.
	\item \emph{Many small problems vs. a single large problem}: The previous contribution is equivalent to the identification of when a CA solving one large problem is more efficient than several agents concurrently solving smaller subproblems.
	\item \emph{Same models for allocation and routing}: In order to focus on the impact of (de-)centralisation only, we try to keep all other things equal. In particular, we propose to minimise the impact of implementation by (i) taking only the computation time of the solver of MILP (Mixed Integer Linear Program) CPLEX into account, (ii) proposing a same allocation MILP for all our decentralised mechanisms, and (iii) relying on close MILP formulations of MTSP and TSP.
	\item \emph{Same structure of state charts for selfish and benevolent agents}: In our previous article, the agents were selfish, whereas this article assumes they are benevolent. It is interesting to note that the structure of the state charts of all our mechanisms remains the same. Only the action carried out in some of the states needs to be modified and more information is exchanged between the agents. Therefore, we do not show these state charts again\footnote{Please refer to \cite{moyaux20} for a presentation of the state charts of our mechanisms. In addition, we will add our AnyLogic model and the outcomes of the experiments to \url{https://github.com/disp-lab/centr_vs_decentr} after this article is accepted for publication.}; instead, we show our mechanisms through their sequence diagram, which is an equivalent way of presenting them. Moreover, two footnotes suggest other mechanisms for benevolent agents with the same structure of state charts.
	\item \emph{Experimental conclusions drawn from two variants of MTSP}: Not only do we show the results for MTSP$_\textrm{b}$, we also summarise the results for MTSP$_\textrm{s}$ reported in our previous article, and note that they are qualitatively different. They are therefore presented as different rankings. The reader does \emph{not need to read our previous article} as we summarise it whenever we cite it.
\end{enumerate}

The outline of this paper is as follows.
Section \ref{sec:mechanisms} describes all our mechanisms.
Section \ref{sec:experiments} details two experiments.
Section \ref{sec:conclusion} concludes.

\section{Our allocation mechanisms to solve the traditional MTSP$_\textrm{b}$}
\label{sec:mechanisms}

This section first introduces definitions and hypotheses before presenting our mechanisms.

%
%
\subsection{Definitions}

\begin{figure*}
	\includegraphics[width=170mm]{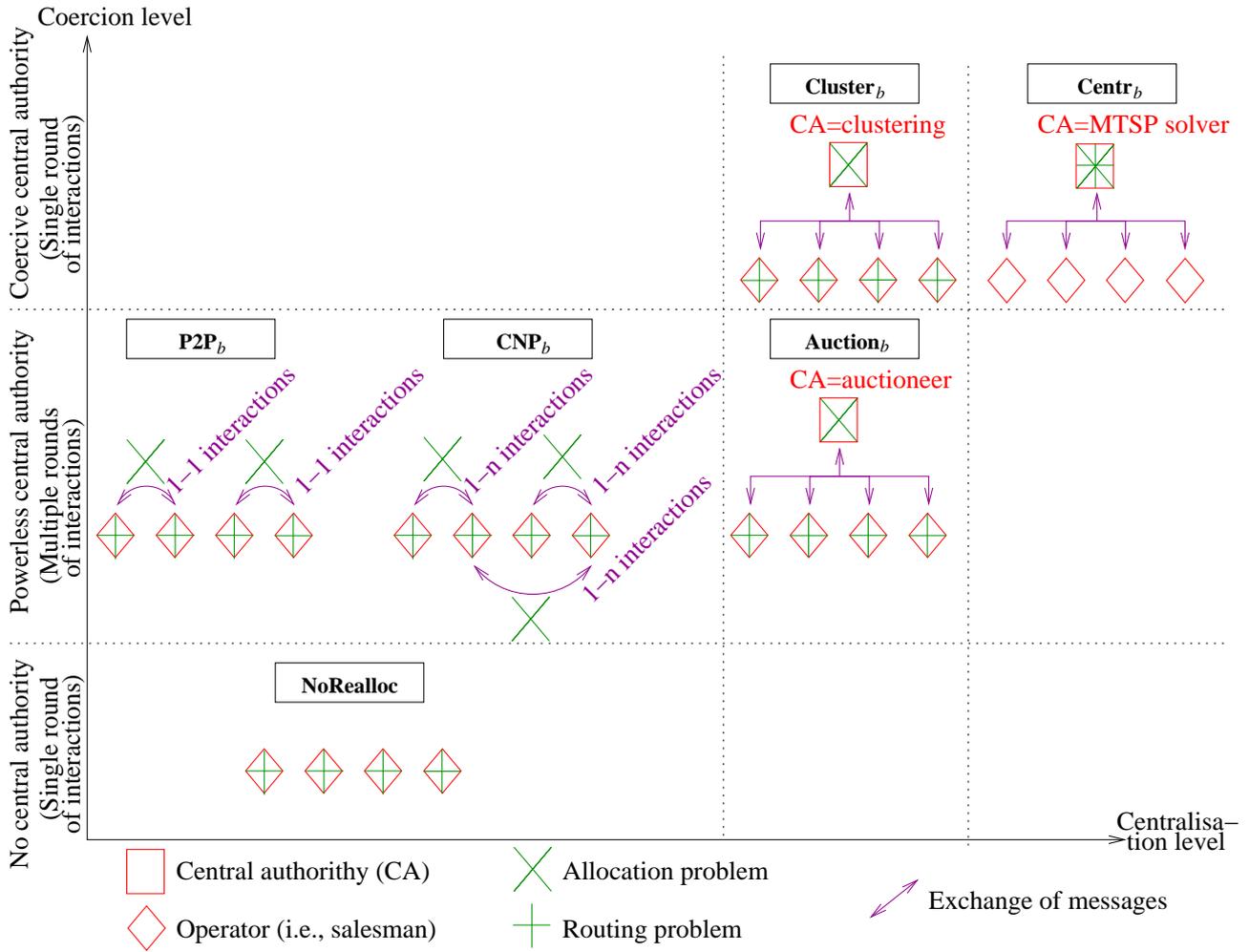}	
	\caption{Overview of the allocations mechanisms.}
	\label{fig:mechanisms}
\end{figure*}

MTSP$_\textrm{b}$ and MTSP$_\textrm{s}$ are the optimisation of two subproblems, namely, the allocation of $n$ cities to $m$ salesmen, and the routing of each salesman such that he visits all the cities allocated to him.
We use ``he'' to designate a salesman who may also be called agent, and ``she'' for the Central Authority (CA), CA is also an agent, but we never need to refer to her this way. We call ``total route length'' the sum of the ``individual route lengths'' of all agents/salesmen.
Figure~\ref{fig:mechanisms} shows the organisations considered in this article.
We call \emph{mechanism} an instance of an organisation, $e.g.,$ Organisation {\sf Cluster$_b$}will be instantiated below as Mechanisms {\sf ClusterR$_b$} and {\sf ClusterS$_b$}.
We can also see in this figure that our organisations may
(i) require a single or multiple rounds of interactions,
(ii) have zero or one CA (who may be either coercive or a facilitator),
(iii) make either the CA or the salesmen solve the routing subproblem,
(iv) etc.
The organisations in Figure~\ref{fig:mechanisms} may be briefly described as:
\begin{itemize}
	\item {\sf NoRealloc} is a benchmark showing the worst/longest total route length by ignoring the allocation subproblem.
	\item {\sf Centr$_b$} finds the best/shortest total route length (when the time limit is sufficient) in a single round by letting CA jointly optimise allocation and routings.
	\item {\sf CNP$_b$} is inspired by the Contract Net Protocol proposed by \cite{smith80}. In each round, one of the salesmen plays the role of the auctioneer of an auction to exchange cities.
	\item {\sf Auction$_s$} also operates by rounds in which the CA plays the role of the auctioneer.
	\item {\sf P2P$_b$} also runs by rounds but only involves two agents who try to exchange cities.
	\item {\sf Cluster$_s$} works in a single round in which all salesmen send their initial endowment of cities to CA, then CA creates clusters of these cities, and the salesmen finally solve a TSP with the cities sent by CA. This organisation has two variants called {\sf ClusterR$_s$} and {\sf ClusterS$_s$} depending on the formulation of the clustering MILP used by CA.
\end{itemize}

We name ``cost'' of a city the additional route length necessary to visit this city, that is, the difference between the lengths of the routes for visiting and not visiting this city. This paragraph details this computation.
During an exchange of cities between two or more agents, we call $c^a$ the city proposed by Agent~$a$, $e.g.,$ $c^{g1}$ and $c^{g1}$ are the cities proposed by guests $g1$ and $g2$ respectively.
We let $d^a_c$ denote the cost for Salesman $a$ when City $c$ is allocated to~$a$.
This cost is the additional distance travelled by $a$ to visit $c$, $i.e.,$ the difference between the length of the solution of the TSP with all the cities already allocated to $a$ plus City $c$, and the length of the solution of the TSP without~$c$.
Similarly, $d^{a1}_{c^{a2},c^{a3}}$ represents the cost for Agent $a1$ to visit Cities $c^{a2}$ and $c^{a3}$ proposed by Agent $a2$ and $a3$. Of course, $d^{a1}_{c^{a2},c^{a3}} = d^{a1}_{c^{a3},c^{a2}}$. The triangle inequality would imply that $d^{a1}_{c^{a2},c^{a3}} \leq d^{a1}_{c^{a2}} + d^{a1}_{c^{a3}}$, but an example below shows that this inequality does not hold.

\emph{Difference with our previous article:} As noted above, we use subscript ``b'' for the mechanisms used in this article, to distinguish them from those with subscript ``s'' in our previous article. Thus, {\sf P2P$_s$}, {\sf CNP$_s$}, {\sf Auction$_s$}, {\sf ClusterR$_s$}, {\sf Centr$_s$} and {\sf Centr$_b$} were previously called {\sf P2P}, {\sf CNP}, {\sf Auction}, {\sf Cluster}, {\sf OptDecentr} and {\sf FullCentr}. {\sf NoRealloc} is the same since subscripts ``b'' and ``s'' are related to allocation and this mechanism ignores this subproblem. {\sf ClusterS$_s$} is not shown because it has results very close to {\sf ClusterR$_s$}.

%
%
\subsection{Hypotheses}

We make an important assumption about the durations taken into account in both our experiments. We choose to consider only the computation time of the solver CPLEX. This is a solution to the problem of comparing, for example, (i) a piece of software written in Java within a month by one person and (ii) a professional library such as CPLEX written in C and improved for decades by a team of developpers. In fact, what we want to study is the impact of (de-)centralisation and we thus try to keep everything else as identical as possible.
This assumption incurs consequences, such as the impossibility to use technniques such as k-means for clustering, the belief–desire–intention software model to represent agents, the neglecting of the travelling time of messsages, etc.

Moreover, we try to make CPLEX solve problems as similar as possible in the various mechanisms. First, we will see that the formulation of TSP in Equations~\ref{eq-TSP-obj}-\ref{eq-TSP-varX}, MTSP in Equations~\ref{eq-MTSP-obj}-\ref{eq-MTSP-varX} and a modified TSP finding the city with the highest cost in Equations~\ref{eq-p2p-guest-obj}-\ref{eq-p2p-guest-varX} are very close.
Next, we introduce the allocation problem in Equations \ref{eq-objective}-\ref{eq-def} such that it can be shared by all our decentralised mechanisms.

Finally, we assume that the agents start with an allocation of cities called inital endowment.


%
%
\subsection{Mechanism {\sf NoRealloc}}

{\sf NoRealloc} is not really a mechanism but a benchmark showing the upper (worst) bound of the total route length when the allocation subproblem is ignored. No exchanges of messages or cities take place and each salesman solves only the TSP in Equations~\ref{eq-TSP-obj}-\ref{eq-TSP-varX} with the cities in his initial endowment.
\begin{eqnarray}
\min	&\sum_{i=0}^{N-1} \sum_{j=0, j\neq i}^{N-1} d_{ij} x_{ij}			\label{eq-TSP-obj}\\
s.t.	&\sum_{j=0, j\neq i}^{N-1} x_{ij}=1		&0\leq i < N			\label{eq-TSP-flowOut}\\
	&\sum_{i=0, i\neq j}^{N-1} x_{ij}=1		&0\leq j < N			\label{eq-TSP-flowIn}\\
	&p_i - p_j + N.x_{ij} \leq N-1		&1\leq i \neq j < N		\label{eq-TSP-subroute}\\
	&p_i\in \Re^+					&1\leq i < N			\label{eq-TSP-varP}\\
	&x_{ij}\in \{0,1\}				&(i,j)\in\{0,1, \ldots N-1\}^2 \label{eq-TSP-varX}
\end{eqnarray}
In this MILP, the binary decision variable $x_{ij}$ in Equation~\ref{eq-TSP-varX} equals 1 only if the considered salesman goes from City $i$ to City~$j$, and zero otherwise. Equation~\ref{eq-TSP-obj} minimises the length of this salesman's individual route, which means that the distance $d_{ij}$ between Cities $i$ and $j$ is added up only when $x_{ij}=1$. This sum is up to $N\not=n$ which is the number of cities currently allocated to the considered salesman (it is not necessary to call it $N_a$ as it is a variable local to Salesman~$a$). Equation~\ref{eq-TSP-flowOut} (respectively, \ref{eq-TSP-flowIn}) ensures that City $i$ (respectively, $j$) is left (respectively, entered) exactly once. The constraint in Equation~\ref{eq-TSP-subroute} eliminates sub-routes by the node potentials method \citep{miller60} which uses the real decision variable $p_i$ in Equation~\ref{eq-TSP-varP} to count the number of cities visited by the considered salesman before he visits City~$i$.

\emph{Difference with our previous article}: {\sf NoRealloc} is the same in both MTSP$_\textrm{b}$ and MTSP$_\textrm{s}$.

%
%
\subsection{Mechanism {\sf Centr$_b$}}

Basically, {\sf Centr$_b$} operates in a single round in which CA solves a well-known MILP formulation of MTSP$_b$ and the salesmen make no decisions.
More precisely, (i) the $m$ salesmen send a message with their initial endowment, (ii) CA jointly solves both allocation and routing subproblems in Equations~\ref{eq-MTSP-obj}-\ref{eq-MTSP-varX}, and (iii) CA sends to each of the $m$ salesmen a message with the cities allocated to them. This list of cities is ordered such that the total route length is minimised.
\begin{eqnarray}
\min 	&\sum_{i=0}^{n-1} \sum_{j=0, j\neq i}^{n-1} d_{ij} x_{ij}		\label{eq-MTSP-obj}\\
	&\sum_{j=1}^{n-1} x_{0j}=m			&			\label{eq-MTSP-flowOutDepot}\\
	&\sum_{j=0, j\neq i}^{n-1} x_{ij}=1		&1\leq i < n		\label{eq-MTSP-flowOut}\\
	&\sum_{i=0, i\neq j}^{n-1} x_{ij}=1		&1\leq j < n		\label{eq-MTSP-flowIn}\\
	&\sum_{i=1}^{n-1} x_{i0}=m			&			\label{eq-MTSP-flowInDepot}\\
	&p_i - p_j + (n-1).x_{ij} \leq n-2		&1\leq i \neq j < n	\label{eq-MTSP-subroute}\\
	&p_i\in \Re^+					&1\leq i < n		\label{eq-MTSP-varP}\\
	&x_{ij}\in \{0,1\}				&0\leq i,j < n		\label{eq-MTSP-varX}
\end{eqnarray}
Equations \ref{eq-MTSP-obj}, \ref{eq-MTSP-subroute}, \ref{eq-MTSP-varP}, \ref{eq-MTSP-varX} are respectively the same as Equations \ref{eq-TSP-obj}, \ref{eq-TSP-subroute}, \ref{eq-TSP-varP}, \ref{eq-TSP-varX}.
Equations \ref{eq-MTSP-flowOut} and \ref{eq-MTSP-flowIn} are respectively the same as Equations \ref{eq-TSP-flowOut} and \ref{eq-TSP-flowIn}, except that they do not apply to the depot, namely City zero which is allocated to all salesmen. Equation \ref{eq-MTSP-flowOutDepot} (respectively, \ref{eq-MTSP-flowInDepot}) is similar to Equation \ref{eq-TSP-flowOut} (respectively, \ref{eq-TSP-flowIn}) but only for the depot, and ensures that $m$ salesmen leaves (respectively, enters) the depot.
As highlighted above, MTSP in Equations~\ref{eq-TSP-obj}-\ref{eq-TSP-varX} and MTSP in Equations~\ref{eq-MTSP-obj}-\ref{eq-MTSP-varX} look very similar, which is a desired feature if we wish to study the impact of (de-)centralisation rather than the consequence of differences of modelling, implementation or anything else.

Note that Equations~\ref{eq-MTSP-flowOutDepot} and \ref{eq-MTSP-flowInDepot} are not inequalities ($\leq$). Both equalities force each of the $m$ salesmen to visit at least one city. The decentralised mechanisms in the next subsections have to take this important detail into account by checking that at least two cities are always allocated to each salesman, $viz.,$ the depot and another city.

\emph{Difference with our previous article}: We have only renamed {\sf OptDecentr} as {\sf Centr$_s$}, and {\sf FullCentr} as {\sf Centr$_b$}.

%
%
\subsection{Mechanism {\sf P2P$_b$}}
\label{sec:P2P}

\begin{figure}
	\centering
	\includegraphics[width=60mm]{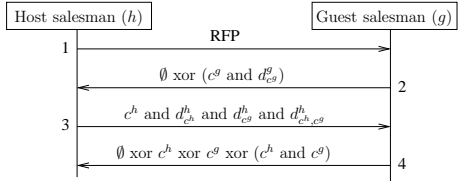}
	\caption{Interactions in every round of {\sf P2P$_b$}.}
	\label{fig:P2P}
\end{figure}

{\sf P2P$_b$} is a mechanism with no CA in which, in every round of interaction, two agents play the role of either a host $h$ or a single guest~$g$.
Figure \ref{fig:P2P} shows the four messages exchanged between $h$ and $g$ in every round.
In this figure, Agents $a$, $a1$ and $a2$ in the above definitions are replaced by $h$ and $g$. For example, $d^g_{c^h}$ is the cost for guest $g$ to visit city $c^h$ which is proposed by the host~$h$. The four messages exchanged in Figure \ref{fig:P2P} may be described as follows:
\begin{enumerate}
	\item The host $h$ sends a RFP (request for participation).
	\item When the guest $g$ is available, he either refuses a round with this specific host by sending $\emptyset$, or accepts by sending both City $c_g$ and his potential reduction of cost $d^g_{c_g}$, $i.e.,$ how long $g$ would reduce his route length by not visiting~$c_g$. City $c_g$ is the city (i) with the highest value of $d^g_{c_g}$ among the cities allocated to $g$, and (ii) which has not been proposed to $h$ yet. Technically, $g$ solves the modified TSP in Equations \ref{eq-p2p-guest-obj}-\ref{eq-p2p-guest-varX} in order to identify~$c_g$.
\begin{eqnarray}
\min	&\sum_{i=0}^{N-1} \sum_{j=0, j\neq i}^{N-1} c_{ij} x_{ij}				\label{eq-p2p-guest-obj}\\
s.t.	&\sum_{j=0, j\neq i}^{N-1} x_{ij}=kept_i	&0\leq i < N				\label{eq-p2p-guest-flowOut}\\
	&\sum_{i=0, i\neq j}^{N-1} x_{ij}=kept_j	&0\leq j < N				\label{eq-p2p-guest-flowIn}\\
	&\sum_{i=1}^{N-1} kept_i = N-2		&0\leq i < N				\label{eq-p2p-guest-nb-of-cities-kept}\\
	&p_i - p_j + N.x_{ij} \leq N-1		&1\leq i \neq j < N			\label{eq-p2p-guest-subroute}\\
	&kept_i = 1					&i|{\tt propCities}[h][i]		\label{eq-p2p-guest-previouslyProposed}\\
	&						&=true	\\
	&kept_i\in \{0,1\}				&i\in\{1, \ldots N-1\}			\label{eq-p2p-guest-var}\\
	&p_i\in \Re^+					&1\leq i < N				\label{eq-p2p-guest-varU}\\
	&x_{ij}\in \{0,1\}				&(i,j)\in\{0,1, \ldots N-1\}^2	\label{eq-p2p-guest-varX}
\end{eqnarray}
Equations \ref{eq-p2p-guest-obj} and \ref{eq-p2p-guest-subroute} are the same as Equations \ref{eq-TSP-obj} and \ref{eq-TSP-subroute}. Equations \ref{eq-p2p-guest-flowOut} and \ref{eq-p2p-guest-flowIn} are similar to Equations \ref{eq-TSP-flowOut} and \ref{eq-TSP-flowIn}, except that the right-hand side does not read 1, but Decision Variable $kept_i$; according to Equation \ref{eq-p2p-guest-nb-of-cities-kept}, $kept_i=1$ for all cities except the one to be proposed to $h$ and which we call $c_g$.
Finally, $g$ memorises in the matrix of Booleans {\tt propCities}$[h][i]$ whether City $i$ has been proposed to this host $h$ in a previous round or not, and Equation \ref{eq-p2p-guest-previouslyProposed} ensures that $c_g$ cannot be City~$i$, $i.e.,$ City~$i$ cannot be proposed to agent $h$ again.

Next, $g$ calculates $d^g_{c_g}$ as the difference between his current route length and the value of the objective function in Equation~\ref{eq-p2p-guest-obj}.
	\item If the interaction has not been stopped by Message 2, $i.e.,$ $g$ did not reply $\emptyset$, then $h$ chooses a City $c_h$ by also solving the modified TSP in Equations~\ref{eq-p2p-guest-obj}-\ref{eq-p2p-guest-varX} (of course, the roles of guest and host need to be swapped in the above explanation of this MILP). Then, $h$ sends $c_h$ and the potential reductions of cost $d^h_{c_h}$, $d^h_{c_g}$ and $d^h_{c_h,c_g}$ to~$g$. An example below illustrates the calculation of these three cost reductions.
	\item Guest $g$ finds the best allocation for all salesmen and either (Case 1) keeps both $c_h$ and $c_g$ by sending $\emptyset$ to $h$, or (Case 2) keeps $c_h$ by sending $c_g$ (the two cities are exchanged), or (Case 3) keeps $c_g$ and returns $c_h$ (no exchange of cities), or (Case 4) keeps nothing and sends both $c_h$ and $c_g$. Since we only take the computation time of CPLEX into account, we need to write the selection of the case incurring the smallest sum of $g$'s and $h$'s costs as a MILP. Simpler formulations are possible for {\sf P2P$_b$} but the following one will also be used in the other mechanisms, since we want to have as much similarity between mechanisms as possible in order to study (de-)centralisation itself. We call:
	\begin{itemize}
	\item $G$: Number of guests; in {\sf P2P$_b$}, $G=1$ (and the $m-G=1$ other salesman is the host).
	\item $A$: Set of participating agents/salesmen; in {\sf P2P$_b$}, $A=\{h,g\}$.
	\item $C$: Set of cities that are proposed in this round of exchange; In {\sf P2P$_b$}, $C=\{c^h,c^g\}$.
	\item $\bar{C}$: Set of bundles of one or two cities that are proposed in this round of exchange. $\bar{C}$ is derived from $C$. In {\sf P2P$_b$}, $\bar{C}=C\cup \{\{c^h, c^g{}\}\} = \{c^h,c^g,\{c^h,c^g\}\}$, $viz.,$ three bundles (two single-city bundles and one two-city bundle).
	\item $C^a$: Set of cities currently allocated to Agent $a$; in any round of {\sf P2P$_b$}, $c^h\in C^h$ and $c^g\in C^g$, $i.e.,$ the city proposed for an exchange by an agent is allocated to this agent in the considered round.
	\item $|X|$: Number of elements in set $X$; for example, in any round of any decentralised mechanism, $|C^a|\geq2$ because all agents $a$ are allocated at least the depot plus one city in order to comply with MTSP$_\textrm{b}$ in Equations \ref{eq-MTSP-obj}-\ref{eq-MTSP-varX}. As another example, in any round of {\sf P2P$_b$}, $|C|=2$ because exactly two agents propose one city each.
	\item $d^a_{\bar{c}}$: Cost (additional distance) for Agent $a$ to visit Bundle $\bar{c}\in \bar{C}$. As noted above, for any two cities $c1$ and $c2$, $d^a_{c1, c2}=d^a_{c2, c1}$, but an example below will show that $d^a_{c1, c2}$ may be either smaller, equal or larger than $d^a_{c1}+d^a_{c2}$, $i.e.,$ this cost does not follow the triangle inequality.
	 In {\sf P2P$_b$}, $[d^A_{\bar{C}}]=\left[ {\begin{array}{cc}
	d^h_{c^h}	&d^g_{c^h}	\\
	d^h_{c^g}	&d^g_{c^g}	\\
	d^h_{c^h,c^g}	&d^g_{c^h,c^g}	\\
\end{array}	}\right]$

	This example shows that $[.]$ is a matrix, $e.g., d^h_{c^h}$ points to one specific value in Matrix $[d^A_{\bar{C}}]$.
	\item $w^c_{\bar{c}}=1$ if Bundle $\bar{c}$ contains city $c$ and $w^c_{\bar{c}}=0$ otherwise.
	For all mechanisms, the first $(1+G)$ rows represent single-city bundles and the last $G$ show two-city bundles.
	In {\sf P2P$_b$}, $[w^C_{\bar{C}}]=\left[ {\begin{array}{cc}
	1	&0\\
	0	&1\\
	1	&1\\
\end{array}	}\right]$

	Again, Matrix $[w^C_{\bar{C}}]$ contains all such values.
	\item $v_{\bar{c}}$ is the number of cities in Bundle $\bar{c}$, $i.e.,$ $v_{\bar{c}}=1$ if $\bar{c}\in C$ ($\bar{c}$ contains a single city) and $v_{\bar{c}}=2$ if $\bar{c}\in \bar{C} \setminus C$ ($\bar{c}$ contains a two-city bundle).
	In {\sf P2P$_b$}, $[v_{\bar{C}}]=\left[ {\begin{array}{c}
	1	\\
	1	\\
	2	\\
\end{array}	}\right]$
	\item $x^a_{\bar{c}}$: This binary decision variable equals one only if Bundle $\bar{c}$ is allocated to agent $a$, and zero otherwise.
	\end{itemize}
Guest $g$ uses CPLEX to solve the allocation MILP in Equations \ref{eq-objective}-\ref{eq-def}.
\begin{eqnarray}
\min		&\sum_{a\in A} \sum_{\bar{c}\in \bar{C}} x^a_{\bar{c}}.d^a_{\bar{c}}		&				\label{eq-objective}\\
{\rm s.t.}	&\sum_{a\in A} \sum_{\bar{c}\in \bar{C}} x^a_{\bar{c}}.w^c_{\bar{c}}=1	&, \forall c\in C		\label{eq-constraint1}\\
		&\sum_{\bar{c}\in \bar{C}} x^a_{c}\leq 1					&, \forall a\in A		\label{eq-constraint2}\\
		&\sum_{\bar{c}\in \bar{C}} x^a_{\bar{c}}.v_{\bar{c}} \geq 3-|C^a|		&, \forall a\in A		\label{eq-constraint3}\\
		&\lefteqn{x^a_{\bar{c}}\in \{0,1\}}						&, \forall a\in A, \bar{c}\in \bar{C}	\label{eq-def}
\end{eqnarray}
The objective function in Equation \ref{eq-objective} minimises the sum of the costs of the salesmen involved in this round of exchange, $e.g., h$ and $g$ in {\sf P2P$_b$}. This minimisation corresponds to a reduction of the total route length travelled by all agents as they are benevolent.
Equation \ref{eq-constraint1} ensures that Bundle(s) $\bar{c}$ is/are selected such that each City $c$ is visited exactly once. Equation \ref{eq-constraint2} prevents the allocation of a pair of cities instead of the two-city bundle containing them, which may occur otherwise because costs $d^a_{\bar{c}}$ do not follow the triangle inequality as illustrated in the example in the next paragraph.
The constraint in Equation \ref{eq-constraint3} ensures that all agents will be allocated at least two cities (the depot plus another city): the left-hand side counts the number of cities in allocation $x^a_{\bar{c}}$ and the right-hand side is the minimal number needed for this agent to keep at least two cities.
\end{enumerate}

\emph{Example}

\begin{figure}
	\centering
	\subfigure[Shortest route for the inital endowment of both agents.]{\includegraphics[width=.49\linewidth]{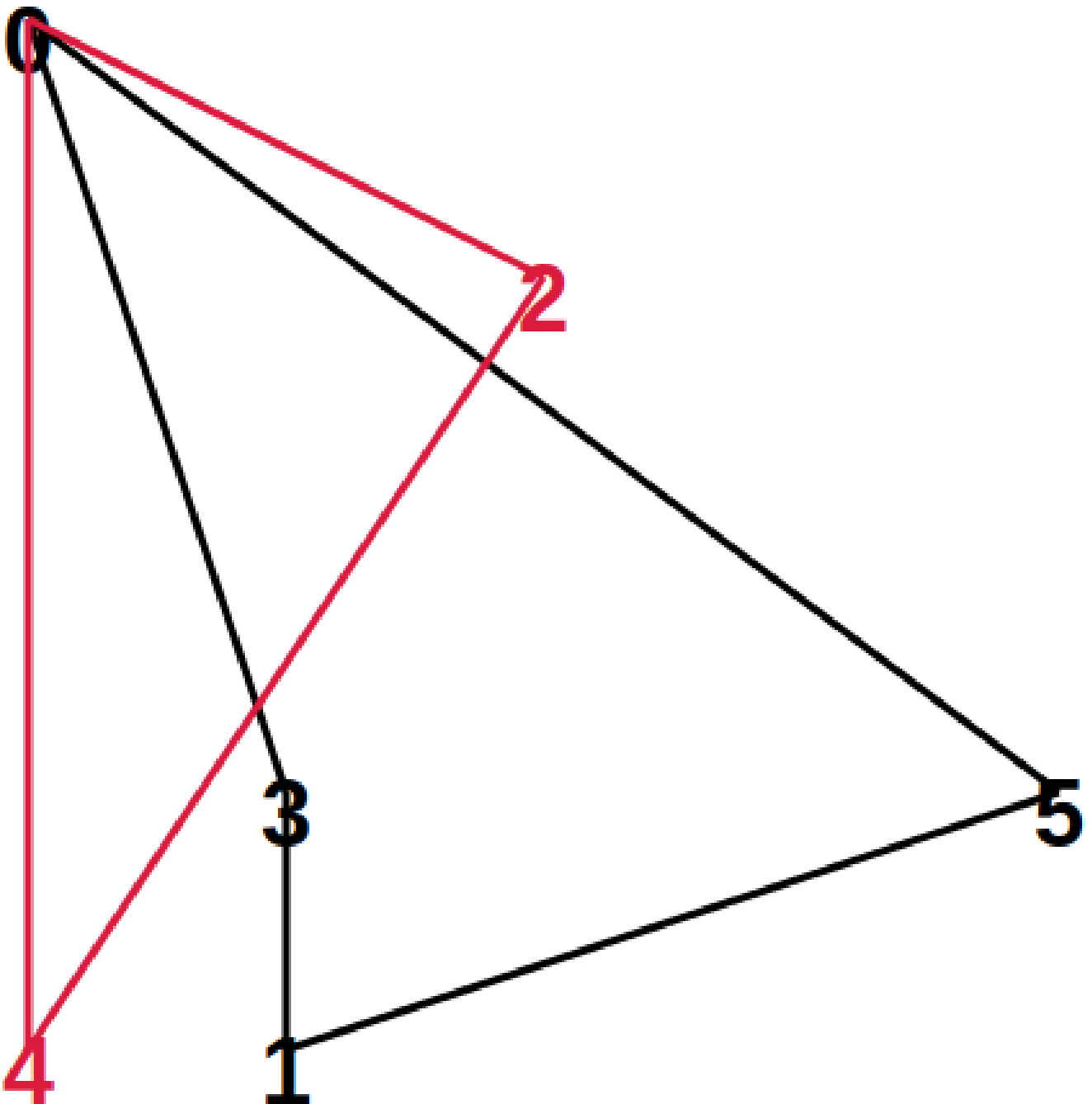}\label{fig:P2P-explanationA}}
	\subfigure[Shortest route for the allocation of both agents after the first round.]{\includegraphics[width=.49\linewidth]{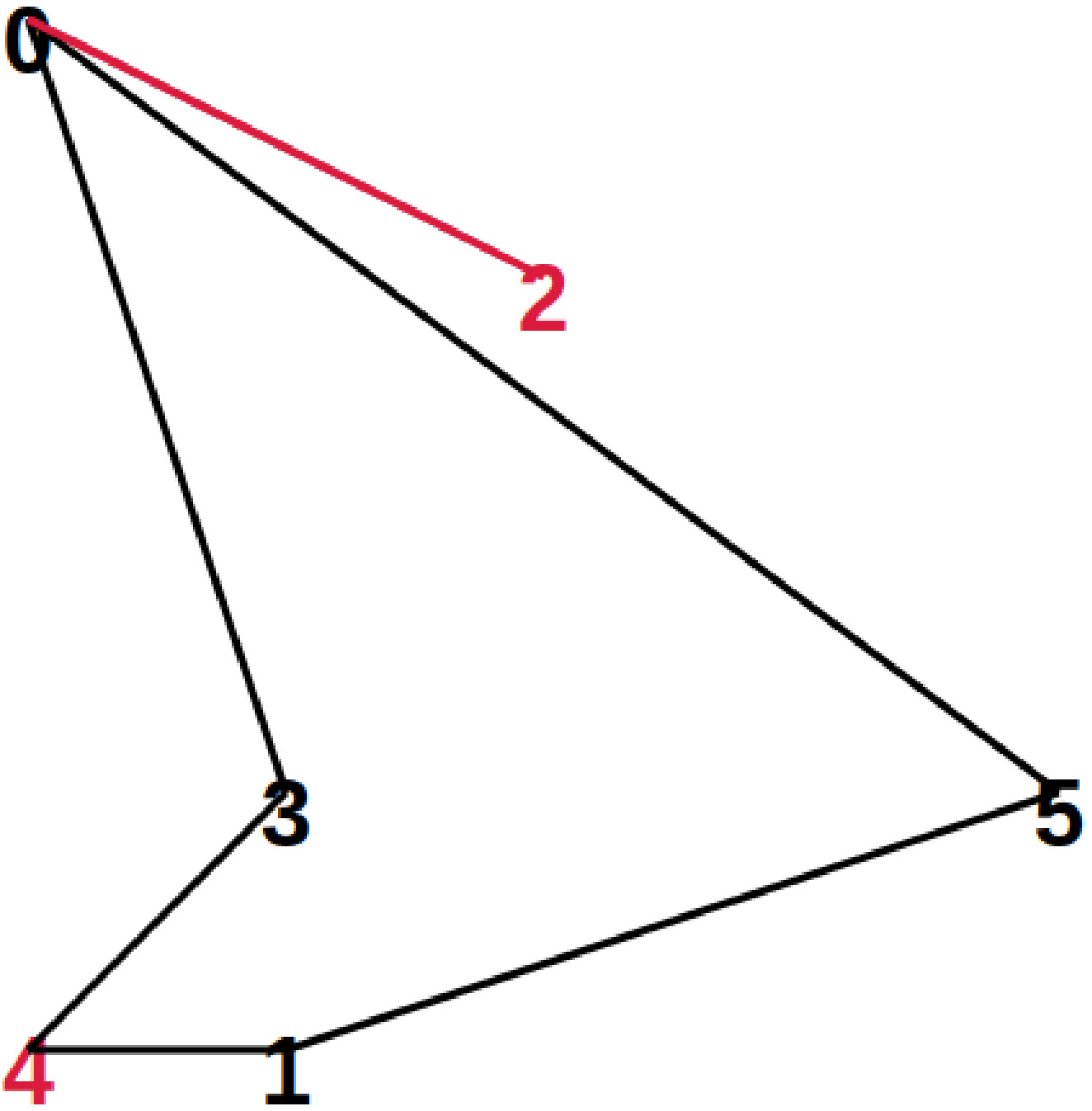}\label{fig:P2P-explanationB}}
	\label{fig:P2P-explanation}
	\caption{Example in AnyLogic showing the need for the constraints in Equation \ref{eq-constraint2} for {\sf P2P$_b$}.}
\end{figure}

We illustrate how the allocation is performed in {\sf P2P$_b$} on an example showing that Matrix $[d^a_{\bar{c}}]$ does not follow the triangle inequality, and why the constraints in Equation~\ref{eq-constraint2} are therefore required. These constraints forbid the allocation of two single-city bundles to the same agent because the corresponding sum of costs would be different from the cost of allocating the bundle of these two cities. This example is shown in Figure~\ref{fig:P2P-explanationA}.
Guest $g$ owns the depot $c_0$ located at (0, 0) and Cities $c_1$ located at (50, 200), $c_3$ at (50, 150) and $c_5$ at (200, 150). Host $h$ is initially allocated the depot $c_0$, as well as Cities $c_2$ at (100, 50) and $c_4$ at (0, 200).
In the first round of interaction, $h$ tries to get rid of $c^h=c_1$, and $g$ of $c^g=c_5$.
$g$ solves the allocation subproblem in Equations \ref{eq-objective}-\ref{eq-def} with $[d^A_{\bar{c}}]=\left[ {\begin{array}{cc}
	d^h_{c^h}	&d^g_{c^h}	\\
	d^h_{c^g}	&d^g_{c^g}	\\
	d^h_{c^h,c^g}	&d^g_{c^h,c^g}	\\
\end{array}	}\right] = \left[ {\begin{array}{ccc}
	268.47	&&43.84	\\
	279.61	&&201.95	\\
	435.77	&&272.66	\\
\end{array}	}\right]$. Let us detail the computation of the right-hand column of this matrix:
\begin{itemize}
	\item $D^g$ is used in the next three bullet points to denote the optimal distance to visit all the cities allocated to $g$ excluding both $c^h$ and $c^g$.
	\item $d^g_{c^h}$ is the difference between (i) the optimal distance to visit all the cities allocated to $g$ including $c^h$ and excluding $c^g$, and (ii) $D^g$.
	\item $d^g_{c^g}$ is the difference between (i) the optimal distance to visit all the cities allocated to $g$ excluding $c^h$ and including $c^g$, and (ii) $D^g$.
	\item $d^g_{c^h,c^g}$ is the difference between (ii) the optimal distance to visit all the cities allocated to $g$ including both $c^h$ and $c^g$, and (ii) $D^g$.
\end{itemize}
Solving the TSP in Equations \ref{eq-TSP-obj}-\ref{eq-TSP-varX} gives the following four optimal routes:
\begin{itemize}
	\item $D^g$ is the length of the route $c_0 \rightarrow c_3 \rightarrow c_1 \rightarrow c_0$, hence $D=\sqrt{(0-50)^2+(0-150)^2} + \\\sqrt{(50-50)^2+(150-200)^2} +\\ \sqrt{(50-0)^2+(200-0)^2}= 414.27$.
	\item $d^g_{c^h}$ is the length of the route $c_0 \rightarrow c_4 \rightarrow c_1 \rightarrow c_3 \rightarrow c_0$ minus $D^g$, hence $d^g_{c^h}=458.11-414.27=43.84$.
	\item $d^g_{c^g}$ is the length of the route $c_0 \rightarrow c_5 \rightarrow c_1 \rightarrow c_3 \rightarrow c_0$ minus $D^g$, hence $d^g_{c^g}=616.23-414.27=201.95$.
	\item $d^g_{c^h,c^g}$ is the length of the route $c_0 \rightarrow c_5 \rightarrow c_1 \rightarrow c_4 \rightarrow c_3 \rightarrow c_0$ minus $D^g$, hence $d^g_{c^h,c^g}=686.94-414.27=272.66$.
\end{itemize}

We have chosen this example because $d^g_{c^h} + d^g_{c^g} = 43.84+201.95 = 245.79 < d^g_{c^h,c^g}=272.66$, that is, costs $d^A_{\bar{c}}$ do not always follow the triangle inequality, even if they sometimes do. Without Equation~\ref{eq-constraint2}, the optimal result of our allocation MILP would be $[x^a_{\bar{c}}]=\left[ {\begin{array}{cc}
	x^h_{c^h}	&x^g_{c^h}	\\
	x^h_{c^g}	&x^g_{c^g}	\\
	x^h_{c^h,c^g}	&x^g_{c^h,c^g}	\\
\end{array}	}\right] = \left[ {\begin{array}{cc}
	0	&1	\\
	0	&1	\\
	0	&0	\\
\end{array}	}\right]$ which would mean that $g$ would keep $c^g$ and add $c^h$ into his allocation. The problem is that this allocation would be found by wrongly assuming a cost of $\sum_{a\in A} \sum_{\bar{c}\in \bar{C}} \left[ {\begin{array}{cc}
	0	&1	\\
	0	&1	\\
	0	&0	\\
\end{array}	}\right].d^a_{\bar{c}}=d^g_{c^h} + d^g_{c^g}=245.79$ only, while this allocation would actually cause the larger increase $\sum_{a\in A} \sum_{\bar{c}\in \bar{C}}$ $\left[ {\begin{array}{cc}
	0	&0	\\
	0	&0	\\
	0	&1	\\
\end{array}	}\right].d^a_{\bar{c}}=d^g_{c^h,c^g} = 276.66$. In other words, visiting two individual cities may have a different cost ($i.e,$ smaller, equal to or larger than) to visiting the bundle of these two cities.
As a result, Equation~\ref{eq-constraint2} forbids this allocation with the contraint $x^g_{c^h} + x^g_{c^g} + x^g_{c^h,c^g} \leq 1$.

\emph{Difference to our previous article}: The allocation of cities in Message 4 in {\sf P2P$_s$} is simpler. It considers only what we call Case 2 ($g$ keeps $c^h$ and $h$ receives $c^g$) and Case 3 (no exchange), and ignores bundles.

%
%
\subsection{Mechanism {\sf CNP$_b$}}

{\sf CNP$_b$} is a decentralised auction in which one of the agents plays the role of host/auctioneer, and the others -- when they agree to take part in the considered round -- are guests.
Figure \ref{fig:CNP} shows the three messages exchanged between a host $h$ and $G\leq (m-1)$ guests (``minus one'' removes the host) called $g1, g2, \ldots gG$. These three messages may be described as follows:
\begin{figure*}[!]
	\centering
	\includegraphics[width=174mm]{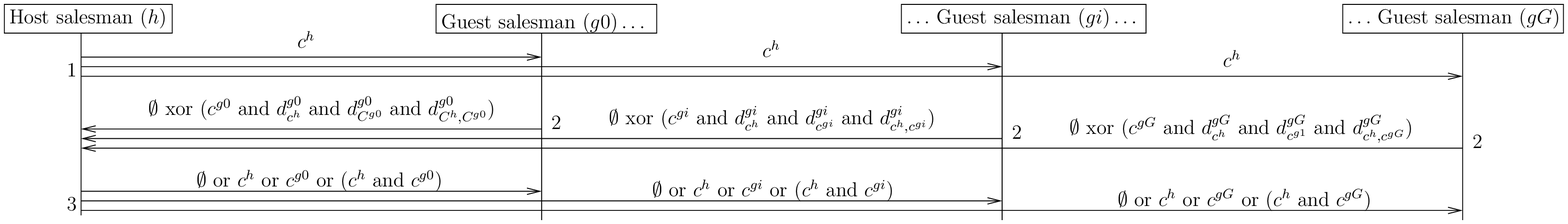}
	\caption{Interactions in every round of {\sf CNP$_b$}.}
	\label{fig:CNP}
\end{figure*}
\begin{enumerate}
	\item One salesman becomes the host $h$ in this round by proposing City $c^h$ to the $(m-1)$ other agents.
	\item Each of the $G\leq (m-1)$ agents interested by $c^h$ becomes Guest $gi$ by replying City $c^{gi}$ and his costs $d^{gi}_{c^h}$, $d^{gi}_{c^{gi}}$ and $d^{gi}_{c^h,c^{gi}}$.
	Like in {\sf P2P$_b$}, $c^{gi}$ is found by solving the modified MTSP in Equations \ref{eq-p2p-guest-obj}-\ref{eq-p2p-guest-varX}.	
	
	The salesmen not interested in $c^h$ reply $\emptyset$ and will be ignored until the start of the next round. To make Figure \ref{fig:CNP} easier to read, we assume that these agents are $g(G+1), \ldots, gm$. They do not appear in the figure.
	\item After the reception of all the cities and costs sent by the guests, $h$ computes the optimal allocation of cities and sends it to the guests. For every guest $gi$, this message contains either zero city ($cf.$ ``$\emptyset$'' in Message 3 in Figure \ref{fig:CNP}), or one city ($cf.$ ``or $c^h$ or $c^{gi}$'' in the figure), or two cities ($cf.$ ``or ($c^h$ and $c^{gi}$)''). For that purpose, $h$ solves the allocation MILP in Equations \ref{eq-objective}-\ref{eq-def}, like in {\sf P2P$_b$} but with the following sets and matrices:
\begin{itemize}
	\item $G\geq 1$ depends on the number of agents not replying $\emptyset$ in Message 2.
	\item $A=\{h, g0, g1, \ldots, g(G-1)$, $gG$\}.
	\item $C=\{c^h, c^{g0}, c^{g1}, \ldots c^{g(G-1)}, c^{gG}\}$.
	\item $\bar{C}=\{c^h, c^{g0}, c^{g1}, \ldots c^{g(G-1)}, c^{gG}, \{c^h,c^{g0}\}, \{c^h,\\c^{g1}\}, \ldots, \{c^h,c^{g(G-1)}\}, \{c^h,c^{gG}\}\}$.
	\item $[d^A_{\bar{c}}]=\\\left[ {\begin{array}{cccccc}
d^h_{c^h}		&d^{g0}_{c^h}		&d^{g1}_{c^h}		&\cdots	&d^{g(G-1)}_{c^h}	&d^{gG}_{c^h}	\\
d^h_{c^{g0}}		&d^{g0}_{c^{g0}}	&\infty		&\cdots	&\infty		&\infty	\\
d^h_{c^{g1}}		&\infty		&d^{g1}_{c^{g1}}	&\cdots	&\infty		&\infty	\\
\vdots			&\vdots		&\vdots		&\ddots	&\vdots		&\vdots\\
d^h_{c^{g(G-1)}}	&\infty		&\infty		&\cdots	&d^{g(G-1)}_{c^{g(G-1)}}	&\infty\\
d^h_{c^{gG}}		&\infty		&\infty		&\cdots	&\infty		&d^{gG}_{c^{gG}}\\
d^h_{c^h,c^{g0}}	&d^{g0}_{c^h,c^{g0}}	&\infty		&\cdots	&\infty		&\infty	\\
d^h_{c^h,c^{g1}}	&\infty		&d^{g1}_{c^h,c^{g1}}	&\cdots	&\infty		&\infty	\\
\vdots			&\vdots		&\vdots		&\ddots	&\vdots		&\vdots\\
d^h_{c^h,c^{g(G-1)}}	&\infty		&\infty		&\cdots	&d^{g(G-1)}_{c^h,c^{g(G-1)}}	&\infty\\
d^h_{c^h,c^{gG}}	&\infty		&\infty		&\cdots	&\infty		&d^{gG}_{c^h,c^{gG}}\\
\end{array}	}\right]$

Clearly, our choice is for guests to compute not all the possible values of $[d^A_{\bar{c}}]$, but only the cost of the two-city bundle containing both $c^h$ and the city proposed by the considered guest. $\infty$ replaces non-calculated costs as these cities should not be given to this agent.\footnote{\label{foot1}We may imagine here two other possible mechanisms for Organisation {\sf CNP$_b$}. The first would make guests compute all values in $[d^A_{\bar{C}}]$ such that $\infty$ would no longer be necessary. The second alternative would consider all two-city bundles in $\bar{C}$ ($e.g., \{c^{g0},c^{g1}\}$ is not considered in our version of the mechanism), which would thus make Matrices $[d^A_{\bar{C}}], [w^C_{\bar{C}}], [v_{\bar{C}}]$ and $[x^A_{\bar{C}}]$ much larger}
In practice, we replace $\infty$ by twice the largest (non-infinite) value in Matrix $[d^A_{\bar{C}}]$.
The values in $[d^A_{\bar{C}}]$ appear in the same order as in $\bar{C}$.

$[d^A_{\bar{C}}]$ has $G+1$ columns and $1+2*G$ rows (``$1$'' for the first line representing $c^h$, plus $G$ rows showing Costs $d^a_c$ for single cities, and the last $G$ rows for distances $d^a_{c^h,c^{gk}}$ for two-city bundles $\{c^h,c^{gk}\}$).
	\item $[w^C_{\bar{C}}]=\left[ {\begin{array}{cccccc}
1			&1			&1			&\cdots	&1		&1	\\
1			&1			&0			&\cdots	&0		&0	\\
1			&0			&1			&\cdots	&0		&0	\\
\vdots			&\vdots		&\vdots		&\ddots	&\vdots	&\vdots\\
1			&0			&0			&\cdots	&1		&0	\\
1			&0			&0			&\cdots	&0		&1	\\
1			&1			&0			&\cdots	&0		&0	\\
1			&0			&1			&\cdots	&0		&0	\\
\vdots			&\vdots		&\vdots		&\ddots	&\vdots	&\vdots\\
1			&0			&0			&\cdots	&1		&0	\\
1			&0			&0			&\cdots	&0		&1	\\
\end{array}	}\right]$

The values in $[w^C_{\bar{C}}]$ appear in the same order as in $\bar{C}$.
\end{itemize}

\end{enumerate}

\emph{Difference with our previous article}: The same messages are exchanged in {\sf CNP$_s$}, but with less information in Message 2. In fact, $gi$ sends ``$\emptyset$ xor ($c^{gi}$ and $d^{gi}_{c^h}$)'' in {\sf CNP$_s$}, instead of ``$\emptyset$ xor ($c^{gi}$ and $d^{gi}_{c^h}$ and $d^{gi}_{c^{gi}}$ and $d^{gi}_{c^{h},c^{gi}}$). Bundles are not considered in {\sf CNP$_s$}: $gi$ sends $\emptyset$ when he does not want City $c^h$, otherwise he sends City $c^{gi}$ and Cost $d^{gi}_{c^h}$ when the exchange of cities would reduce his personnal route length. In the latter case, $h$ accepts the exchange only if this reduces his individual route length as well.

%
%
\subsection{Mechanism {\sf Auction$_b$}}

\begin{figure*}[!]
	\centering
	\includegraphics[width=174mm]{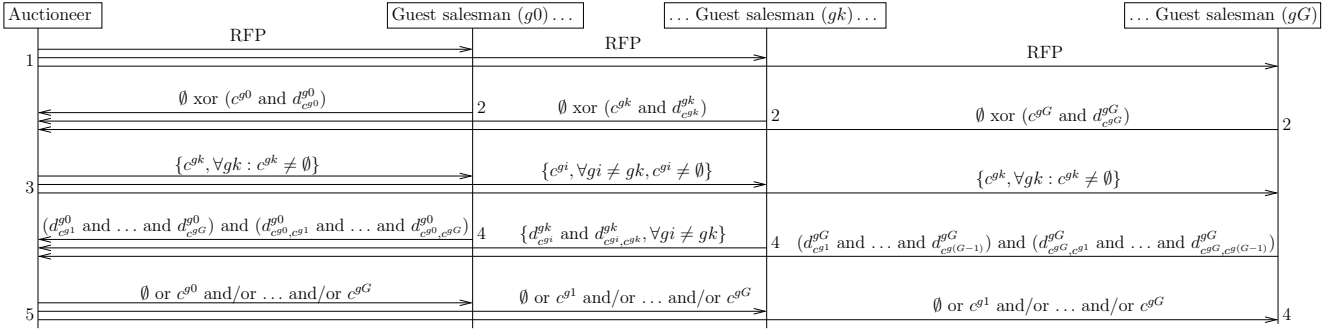}
	\caption{Interactions in every round of {\sf Auction$_b$}.}
	\label{fig:auction}
\end{figure*}

{\sf Auction$_b$} is similar to {\sf CNP$_b$}, except that the role of the auctioneer is not played by one of the salesmen, but by CA.
Figure~\ref{fig:auction} shows the interactions in one of several rounds of interaction.
To make this figure easier to read, we assume again that agents $g0, \ldots, gG$ agree to participate in this round by not replying $\emptyset$ in Message 2, and the other $(m-G)$ agents reply~$\emptyset$ and are hence ignored until the start of the next round.

\begin{itemize}
	\item $A=\{g0, g1, g2, \ldots, g(G-1)$, $gG$\}.
	\item $C=\{c^{g0}, c^{g1}, c^{g2}, \ldots c^{g(G-1)}, c^{gG}\}$.
	\item $\bar{C}=C\cup \{\{c^{gi}, c^{gj}\}, \forall i<j\}$, hence:\\
	$\bar{C}=\{c^{g0}, c^{g1}, c^{g2},\ldots c^{g(G-1)}, c^{gG},
	\{c^{g0},c^{g1}\}, \{c^{g0},c^{g2}\},\\ \ldots \{c^{g0},c^{g(G-1)}\}, \{c^{g0},c^{gG}\},
	\{c^{g1},c^{g2}\}, \{c^{g1},c^{g3}\}, \ldots, \\ \{c^{g1},c^{g(G-1)}\}, \{c^{g1},c^{gG}\}, \ldots,
	\{c^{g(G-3)},c^{g(G-2)}\},\\ \{c^{g(G-3)},c^{g(G-1)}\}, \{c^{g(G-3)},c^{gG)}\}, 
	\{c^{g(G-2)},c^{g(G-1)}\},\\ \{c^{g(G-2)},c^{gG)}, \{c^{g(G-1)},c^G\}	\}$.
	
	The first $G$ rows in this vector represent single cities, and the other $G*(G-1)/2$ rows represent two-city bundles, hence, $\bar{C}$ has $G+G*(G-1)/2$ rows. Let us explain the reason for $G*(G-1)/2$ rows: It is easy to check that there are 1 two-city bundle when there are $G=2$ guests, 3 bundles when $G=3$, 6 when $G=4$, 10 when $G=5$, and so forth, and $G*(G-1)/2$ two-city bundles for $G$ guests.
	That is, for $G$ guests, the two-city bundles are the same as for $(G-1)$ guests, plus $G$ additional two-city bundles containing guest $gG$'s City~$c^{gG}$. Hence, the number of two-city bundles equals the sum of all integers between 1 and $G$.
	\item $[d^A_{\bar{C}}]=\\ \small \left[ {\begin{array}{ccccc}
d^{g0}_{c^{g0}}		&d^{g1}_{c^{g0}}	&\cdots	&d^{g(G-1)}_{c^{g0}}		&d^{gG}_{c^{g0}}	\\
d^{g0}_{c^{g1}}		&d^{g1}_{c^{g1}}	&\cdots	&d^{g(G-1)}_{c^{g1}}		&d^{gG}_{c^{g1}}	\\
d^{g0}_{c^{g2}}		&d^{g1}_{c^{g2}}	&\cdots	&d^{g(G-1)}_{c^{g2}}		&d^{gG}_{c^{g2}}	\\
\vdots				&\vdots		&\ddots	&\vdots			&\vdots\\
d^{g0}_{c^{g(G-1)}}		&d^{g1}_{c^{g(G-1)}}	&\cdots	&d^{g(G-1)}_{c^{g(G-1)}}	&d^{gG}_{c^{g(G-1)}}\\
d^{g0}_{c^{gG}}		&d^{g1}_{c^{gG}}	&\cdots	&d^{g(G-1)}_{c^{gG}}		&d^{gG}_{c^{gG}}\\

d^{g0}_{c^{g0},c^{g1}}		&d^{g1}_{c^{g0},c^{g1}}&\cdots	&\infty			&\infty	\\
d^{g0}_{c^{g0},c^{g2}}		&\infty		&\cdots	&\infty			&\infty	\\
\vdots				&\vdots		&\ddots	&\vdots			&\vdots	\\
d^{g0}_{c^{g0},c^{g(G-1)}}	&\infty		&\cdots	&d^{g(G-1)}_{c^{g0},c^{g(G-1)}}&\infty\\
d^{g0}_{c^{g0},c^{gG}}		&\infty		&\cdots	&\infty			&d^{gG}_{c^{g0},c^{gG}}\\
\infty				&d^{g1}_{c^{g1},c^{g2}}&\cdots	&\infty			&\infty	\\
\infty				&d^{g1}_{c^{g1},c^{g3}}&\cdots	&\infty			&\infty	\\
\vdots				&\vdots		&\ddots	&\vdots			&\vdots	\\
\infty				&d^{g1}_{c^{g1},c^{g(G-1)}}		&\cdots	&d^{g(G-1)}_{c^{g1},c^{g(G-1)}}&\infty\\
\infty				&d^{g1}_{c^{g1},c^{gG}}&\cdots	&\infty			&d^{gG}_{c^{g1},c^{gG}}\\
\vdots				&\vdots		&\ddots	&\vdots			&\vdots	\\
\infty				&\infty		&\cdots	&d^{g(G-1)}_{c^{g(G-3)},c^{g(G-1)}}	&\infty	\\
\infty				&\infty		&\cdots	&\infty			&d^{g(G-1)}_{c^{g(G-3)},c^{gG}}\\
\infty				&\infty 		&\cdots	&d^{g(G-1)}_{c^{g(G-2)},c^{g(G-1)}}	&\infty	\\
\infty				&\infty		&\cdots	&\infty			&d^{gG}_{c^{g(G-2)},c^{gG}}\\
\infty				&\infty		&\cdots	&d^{g(G-1)}_{c^{g(G-1)},c^{gG}}&d^{gG}_{c^{g(G-1)},c^{gG}}\\
\end{array}	}\right]$

	Like $\bar{C}$, $[d^A_{\bar{C}}]$ has $G+G*(G-1)/2$ rows. This matrix also has $G$ columns.
	
	In our version of {\sf Auction$_b$}, each guest $gk$ sends ``\{$d^{gk}_{c^{gi}}$ and $d^{gk}_{c^{gi},c^{gk}}, \forall gi \not= gk$\}'' in Message 4, that is, he proposes a cost for all other agent's one-city bundles, but not for all their two-city bundles, $i.e.,$ $gk$ only proposes a cost for two-city bundles in which one of both cities is $c^{gk}$. Like in {\sf CNP$_b$}, the missing costs in $[d^a_{\bar{c}}]$ are replaced by $\infty$.\footnote{\label{foot2}We may imagine other mechanisms for Organisation {\sf Auction$_b$}. Basically, the guests may compute and send the costs of all two-city bundles in Message 4, such that $[d^a_{\bar{c}}]$ would contain all costs and no $\infty$. Next, we may consider three-city bundles, four-city bundles, $\ldots$, and even all the possible bundles which may be formed with the cities submitted in Message 2.}

	\item $[w^c_{\bar{c}}]=\left[ {\begin{array}{cccccccc}
1	&0		&0		&\cdots	&0	&0		&0	\\
0	&1		&0		&\cdots	&0	&0		&0	\\
0	&0		&1		&\cdots	&0	&0		&0	\\
\vdots	&\vdots	&\vdots	&\ddots	&\vdots &\vdots	&\vdots\\
0	&0		&0		&\cdots	&0	&1		&0	\\
0	&0		&0		&\cdots	&0	&0		&1	\\
1	&1		&0		&\cdots	&0	&0		&0	\\
1	&0		&1		&\cdots	&0	&0		&0	\\
\vdots	&\vdots	&\vdots	&\ddots	&\vdots&\vdots		&\vdots\\
1	&0		&0		&\cdots	&1	&0		&0	\\
1	&0		&0		&\cdots	&0	&1		&0	\\
1	&1		&0		&\cdots	&0	&0		&1	\\
0	&1		&0		&\cdots	&0	&0		&0	\\
\vdots	&\vdots	&\vdots	&\ddots	&\vdots&\vdots		&\vdots\\
0	&1		&0		&\cdots	&0	&1		&0	\\
0	&1		&0		&\cdots	&0	&0		&1	\\
\vdots	&\vdots	&\vdots	&\ddots	&\vdots &\vdots	&\vdots\\
0	&0		&0		&\cdots	&0	&1		&0	\\
0	&0		&0		&\cdots	&0	&0		&1	\\
0	&0		&0		&\cdots	&0	&1		&0	\\
0	&0		&0		&\cdots	&0	&0		&1	\\
0	&0		&0		&\cdots	&0	&1		&1
\end{array}	}\right]$

\end{itemize}

\emph{Difference compared to our previous article}: {\sf Auction$_s$} exchanges the same messages with the same contents as {\sf Auction$_b$}, except Message 4 which has less information. This message  for Salesman $gk$ in {\sf Auction$_s$} is ``\{$d^{gk}_{c^{gi}}, \forall gi \not= gk$\}'', instead of ``\{$d^{gk}_{c^{gi}}$ and $d^{gk}_{c^{gi},c^{gk}}, \forall gi \not= gk$\}'' in Figure~\ref{fig:auction}, $i.e.,$ bundles are not taken into account in the selfish version.

%
%
\subsection{Mechanisms {\sf Cluster$_b$}: {\sf ClusterR$_b$} and {\sf ClusterS$_b$}}

In Organisation {\sf Cluster$_b$}, (i) the $m$ salesmen send all their cities to CA, then (ii) CA creates $m$ clusters with these $n$ cities and sends this allocation back to each salesman who, eventually, (iii) locally solves the TSP in Equations \ref{eq-TSP-obj}-\ref{eq-TSP-varX} with the cities allocated to him.
We consider two MILP used by CA to create these $m$ clusters: (i) {\sf ClusterR$_b$} relies on the formulation proposed by \cite{rao71}, while (ii) {\sf ClusterS$_b$} relies on a variant of this MILP proposed by \cite{saglam06}.

\emph{Difference compared to our previous article}: {\sf ClusterR$_b$} uses the orignal formulation by \cite{rao71}, and {\sf ClusterS$_b$} the model by \cite{saglam06}, while their selfish versions have one additional constraint in order to model the selfishness of agents. {\sf ClusterR$_s$} was called {\sf Cluster} in our previous article, and we only mentionned {\sf ClusterS$_S$} to say that its performance is always very similar to {\sf ClusterR$_S$}. By contrast, in Subsection \ref{sec:exp2} we show that {\sf ClusterR$_b$} and {\sf ClusterS$_b$} may have different results for short period of time.

\section{Experimental comparisons}
\label{sec:experiments}

This section shows two experiments.
Subsection \ref{sec:exp1} reproduces the same experiment as in our previous article, to establish whether the results are the same as in the ``benevolent'' version of our mechanisms. We see that this is not the case, and these experiments thus show that the level of (de-)centralisation is not the only determinant of efficiency.
Subsection \ref{sec:exp2} presents a second experiment investigating when (de-)centralisation is most efficient. For this purpose, we gradually reduce the period of computation time.

We call ``instance'' the location of the $n$ cities and their initial allocation to the $m$ salesmen. The instances are different in both experiments, to check that they have no impact on the results. Both sets use the Euclidean distance.

\subsection{Experiment 1 : Study of the impact of the level of (de-)centralisation on efficiency}
\label{sec:exp1}

The first experiment addresses the following question: if a mechanism $m1_\textrm{s}$ for MTSP$_\textrm{s}$ is better than another $m2_\textrm{s}$, then is its ``benevolent'' version $m1_\textrm{b}$ also better than $m2_\textrm{s}$? 
In other words, is the level of (de-)centralisation the only feature impacting performance?
We explore this question by comparing the results of the mechanisms detailed in this article with those in our previous paper.\footnote{As noted above, this article is complete within itself; accordingly, our older results are described.}
We see that the answer is negative, because the ranking of mechanisms is different in MTSP$_\textrm{b}$ and MTSP$_\textrm{s}$.

\subsubsection{Settings}

In the first experiment, the instances with $m=9$ salesmen are the same as in our previous article: Each city is located at the same place and allocated to the same salesman (same initial endowment), as detailed below. Conversely to our previous work which also showed the results for $m=5$ salesmen, we think it more informative to show the results for $m=2$ instead, so that the difference between these two sizes of instances is bigger.
We compare our mechanisms on the same samples of 130 instances generated by the circular permutation of the ordinate of the 130 cities in ``CH130'' from TSPLIB (\url{http://comopt.ifi.uni-heidelberg.de/software/TSPLIB95/tsp/ch130.tsp.gz}) and use only the first $n$ of these cities. That is, the $c^\textrm{th}$ city in our $i^\textrm{th}$ instance has the same abscissa as the $c^\textrm{th}$ city in CH130, and the same ordinate as the $((c+i)\%130)^\textrm{th}$ city in CH130 (where ``\%130'' represents the rest of the division by 130).
The initial endowment is also obtained by circular permutation: (City 0 is the depot allocated to all salesmen, and) City 1 is allocated to Salesman 0, City 2 to Salesman 1, $\ldots$, City $m$ to Salesman $(m-1)$, City $(m+1)$ to Salesman 0, City $(m+2)$ to Salesman 1, and so on until City $(n-1)$.
As a consequence, the $c^\textrm{th}$ city in the $i^\textrm{th}$ instance with $n$ cities has the same location and is initially allocated to the same salesman as the $c^\textrm{th}$ city in the $i^\textrm{th}$ instance with $(n+1)$ cities and, of course, the latter instance has one additional city.

\begin{figure*}
	\begin{center}
	\includegraphics[width=8.5cm,height=10.7cm]{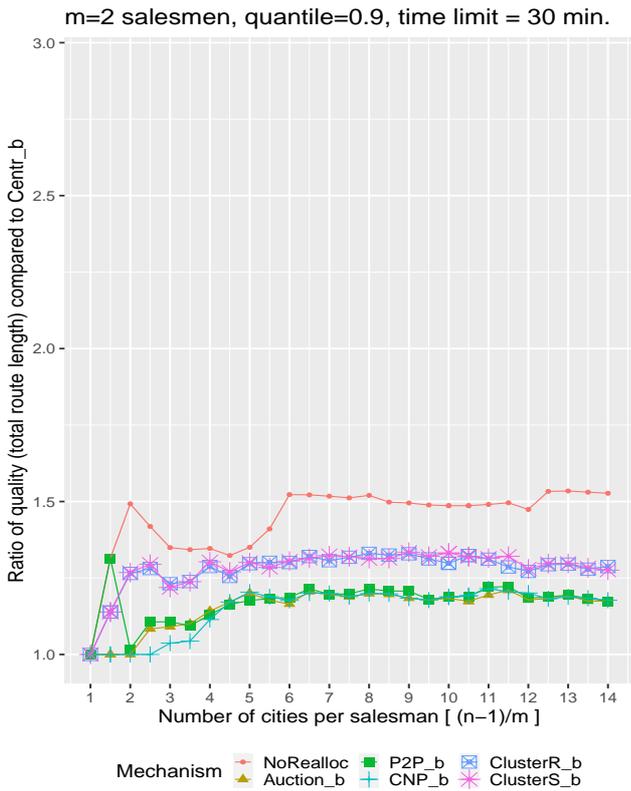}
	\includegraphics[width=8.5cm,height=10.7cm]{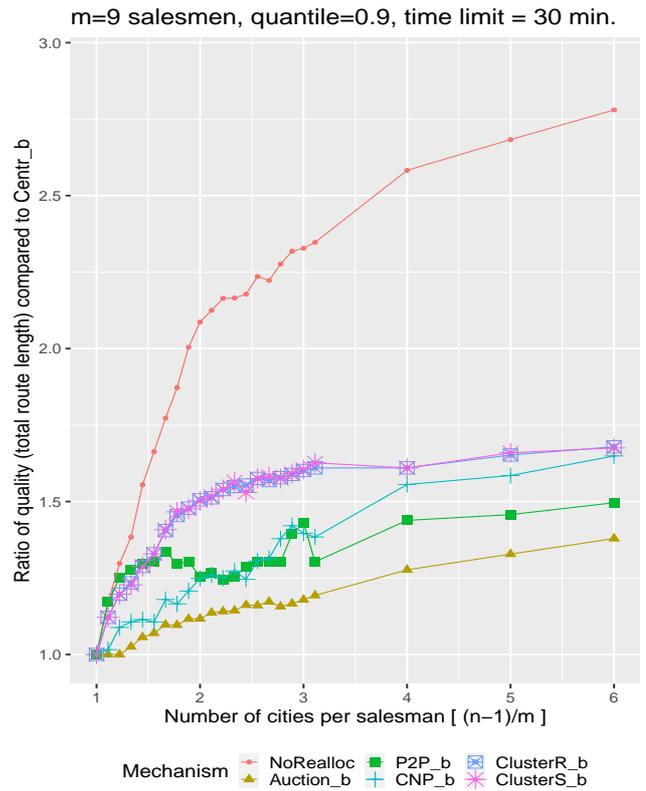}
	\includegraphics[width=8.5cm,height=10.7cm]{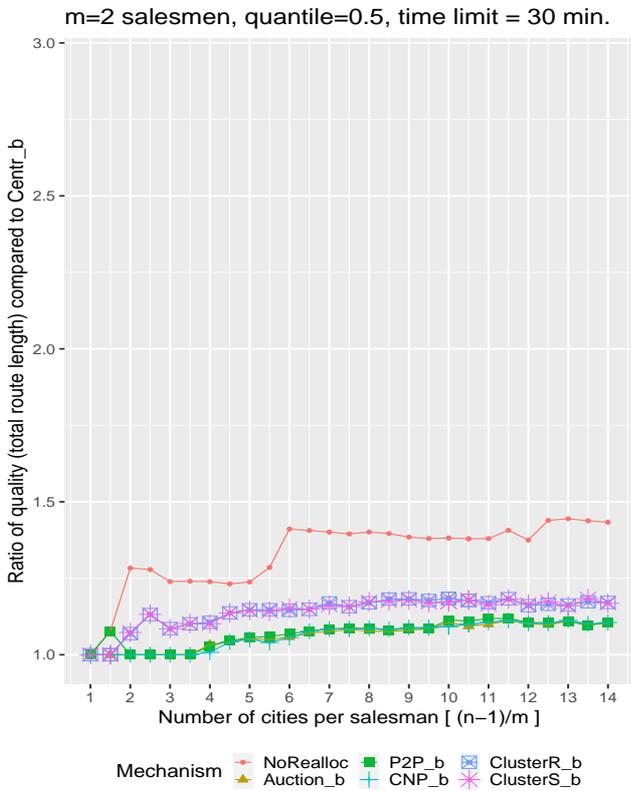}
	\includegraphics[width=8.5cm,height=10.7cm]{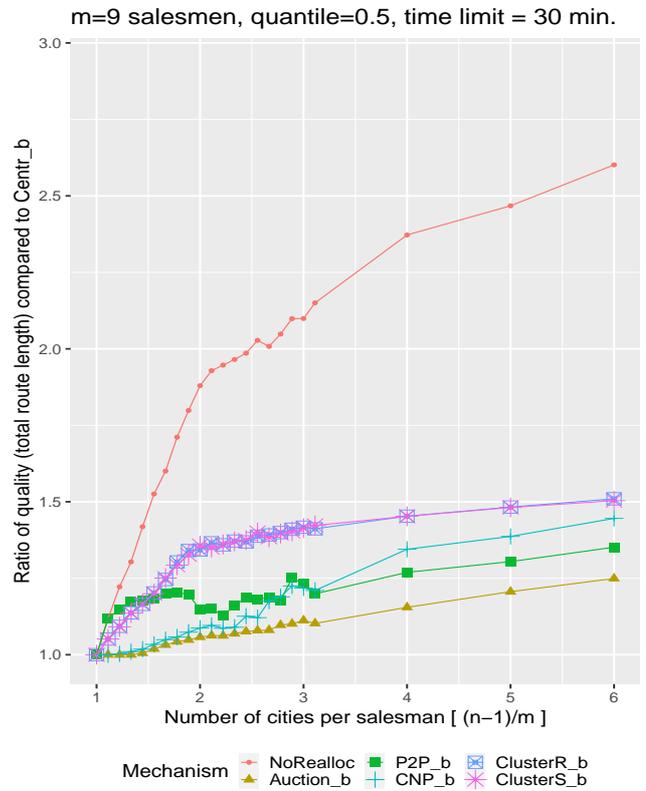}
	\caption{Experiment 1: Deciles of the ratios of quality of every mechanism compared to Mechanism {\sf Centr$_b$} for $m=2$ (left) and $m=9$ (right) salesmen and $n$n cities generated from TSPLIB-CH130 with a time limit of 30 minutes.}
	\label{fig-ratios-30min-centrb}
	\end{center}
\end{figure*}

Each point in Figure \ref{fig-ratios-30min-centrb} is the ninth decile (top) or median (bottom) over these 130 instances of the ratios of the route length found by the considered mechanism in less than 30 minutes of computation divided by the length found by Mechanism {\sf Centr$_b$} in the same time span.
Shortly, \emph{the 130 instances are solved seven times (once per mechanism), then the total route length found by one mechanism is divided by the length found by {\sf Centr$_b$} and, finally, a decile of these 130 ratios is plotted as a point} in Figure~\ref{fig-ratios-30min-centrb}.

The time limit of 30 minutes takes into account concurrency. For instance, each of the two salesmen in {\sf P2P$_b$} may compute for up to 30 minutes. As another example, in {\sf CNP$_b$}, if CA needs 15 minutes to create clusters, then each salesman has up to 15 minutes to locally solve the TSP in Equations \ref{eq-p2p-guest-obj}-\ref{eq-p2p-guest-varX}. Section 4 and Subsection 5.3 in our previous article detail this point and everything is exactly the same here.

The results in this subsection were obtained on PCs with Intel Core i7-10700 CPU @ 2.90 GHz and 16 GB RAM on 64-bit Windows 10 running AnyLogic 8.7.2 (configured to use 4,096 Mb of memory) and CPLEX 12.10.

\subsubsection{Results}

The results in Figure \ref{fig-ratios-30min-centrb} show that the best mechanism is always {\sf Centr$_b$} because all the (ninth deciles and medians of the) ratios comparing every mechanism to {\sf Centr$_b$} are larger than 1 for all the considered sizes of instances. To be precise, for larger instances and $m=2$ salesmen, the median (respectively, the ninth decile) of the total route lengths of {\sf Auction$_b$}, {\sf CNP$_b$} and {\sf P2P$_b$} reaches a plateau $\approx$10\% (respectively, $\approx$20\%) longer than {\sf Centr$_b$}. In the same way, the medians of {\sf ClusterR$_b$} and {\sf ClusterS$_b$} are $\approx$17\% worse (longer) than {\sf Centr$_b$} for $m=2$, and their ninth deciles are $\approx$30\% worse. The medians of {\sf NoRealloc} are $\approx$40\% worse and its ninth deciles $\approx$50\% worse.
However, such a plateau has not been reached yet for $m=9$, but we can see for 6 cities per salesman ($i.e., n=55$ cities) that the medians of 
{\sf Auction$_b$}, {\sf P2P$_b$}, {\sf CNP$_b$}, {\sf ClusterR$_b$}/{\sf ClusterS$_b$} and {\sf NoRealloc} are 
$25\%$, $35\%$, $45\%$, $45\%$ and $60\%$ worse respectively than {\sf Centr$_b$}; The corresponding ninth deciles are
$38\%$, $50\%$, $65\%$, $68\%$ and $78\%$ worse.

We may conclude from these numbers that, conversely to our previous article, \emph{our results no longer show that more centralisation induces more efficiency} because, for example, {\sf P2P$_b$} is at the same time more decentralised and more efficient than {\sf CNP$_b$}, {\sf ClusterR$_b$} and {\sf ClusterS$_b$}.

Another way to state this result is to note that the rankings ($i.e.,$ relative efficiencies) of our mechanisms are not the same with their ``benevolent'' and ``selfish'' versions.
Table \ref{tab:comparison} summarises the rankings in our previous paper, in this experiment and that in the next subsection.
\begin{table}
	\centering
	\begin{tabular}{c||c|c}
				&\multicolumn{2}{c}{Time limit} 			\\
	Variant		&too short for {\sf Centr}	&long enough for {\sf Centr}	\\
	of MTSP		&to obtain optimum		&to obtain optimum	\\
	\hline	\hline
				&{\sf ClusterR$_b$}		&{\sf Centr$_b$}	\\
				&$>$ {\sf Auction$_b$}		&$\gg$ {\sf Auction$_b$}	\\
				&$\approx$ {\sf P2P$_b$}	&$>$ {\sf P2P$_b$} \\
				&$>$ {\sf CNP$_b$}  		&$>$ {\sf CNP$_b$} \\
	MTSP$_\textrm{b}$	&$>$ {\sf ClusterS$_b$} 	&$>$ {\sf ClusterS$_b$} \\
				&				&$\approx$ {\sf ClusterR$_b$}	\\
				&$\gg$ {\sf NoRealloc}		&$\gg$ {\sf NoRealloc}		\\
				&$\gg$ {\sf Centr$_b$}		&				\\
				&				&[all $n$ in Fig.~\ref{fig-ratios-30min-centrb}]	\\
				&[$n=121$ and 			&and [$n=121$ and 		\\
				&time limit $<7$ s.		&time limit $>28$ s.	\\
				&in median in Fig. \ref{fig-varying-time}]	&in Fig. \ref{fig-varying-time}]	\\
	\hline
				&				&{\sf Centr$_s$}	\\
				&{\sf ClusterR$_s$}		&$\gg$ {\sf ClusterR$_s$}	\\
				&$\approx$ {\sf ClusterS$_s$}	&$\approx$ {\sf ClusterS$_s$}	\\
				&$\gg$ {\sf CNP$_s$}		&$\gg$ {\sf CNP$_s$} 		\\
	MTSP$_\textrm{s}$	&$\approx$ {\sf Auction$_s$}	&$\approx$ {\sf Auction$_s$}	\\
				&$>$ {\sf P2P$_s$}		&$>$ {\sf P2P$_s$} 	\\
				&$\gg$ {\sf Centr$_s$}		&			\\
				&$\gg$ {\sf NoRealloc}		&$\gg$ {\sf NoRealloc}		\\
				&[$n>46$ and time		&[$n<37$ and time	\\
				&limit=1,800 s. in		&limit=1,800 s. in	\\
				&Fig. 9 in Moyaux		&Fig. 9 in Moyaux	\\
				&and Marcon (2020)]		&and Marcon (2020)]
	\end{tabular}
	\caption{Ranking of the mechanisms for $m=9$ salesmen ($\gg$ means ``is a much better mechanism than'', $>$ ``is better than'' and $\approx$ ``is as good as'')}
	\label{tab:comparison}
\end{table}
What we have just observed in Figure \ref{fig-ratios-30min-centrb} is summarised on the right-hand side of the first row in this table: {\sf Centr$_b$} performs much better than {\sf Auction$_b$}, which performs slightly better than {\sf P2P$_b$}, $\ldots$, {\sf ClusterR$_b$} and {\sf ClusterS$_b$} are equivalent, and {\sf NoRealloc} has the worst performance.
The next subsection will obtain the same ranking for other instances when the time limit is sufficient for {\sf Centr$_b$} to find the optimum (over 28 seconds).

By contrast, the right-hand side of the second row in Table \ref{tab:comparison} summarises the fact that we had another ranking with the ``selfish'' version of our mechanisms. This ranking shows that more centralisation induces better efficiency for MTSP$_\textrm{s}$. In other words, our older results for MTSP$_\textrm{s}$ support the widespread intuition that centralisation is best when there is no issue of reactivity.

If we compare the two columns in the second row in Table \ref{tab:comparison}, we may conclude that adding a constraint modelling the selfishness of agents may make {\sf Centr$_s$} the best mechanism for small instances (less than 5 cities per salesman), and the worst for larger instances when the time frame is too short to find the optimal solution.

We have just had a look at three of the four entries in Table \ref{tab:comparison}. The left-hand side of its first row shows the rankings when {\sf Centr$_b$} does not have enough time to find the optimal solution. The next subsection shows where this information comes from.

\subsection{Experiment 2 : Search for the time limit favouring (de-)centralisation}
\label{sec:exp2}

The experiment presented in this subsection addresses the following question: when does decentralisation outperform centralisation with our mechanisms for MTSP$_\textrm{b}$? Or, more technically, below what time limit do {\sf P2P$_b$}, {\sf CNP$_b$}, {\sf Auction$_b$} and/or {\sf Cluster$_b$} outperform {\sf Centr$_b$}?


For instances small enough for {\sf Centr$_b$} to complete its optimisation and hence find the best (shortest) route, decentralised mechanisms cannot find a better solution as it is the optimal one. In our previous paper, we saw that all the tested decentralised mechanisms found better solutions than {\sf Centr$_s$} when the time limit prevents {\sf Centr$_s$} from completing its optimisation, as just seen on the left-hand side of the second row in Table \ref{tab:comparison}.
Hence, the following experiments gradually modify one of the three interconnected levers which may make (de-)centralisa\-tion the best option:
\begin{itemize}
	\item \emph{$m$ and $n$}: These two parameters change the size of the instances. The larger $n$, the larger the computation time of all mechanisms. For a given $n$, increasing $m$ favours decentralisation because the salesmen locally solve their TSP with fewer cities.
	\item \emph{Time limit}: Since our first experiment sets the maximum duration of computation at 30 minutes and makes $m$ and $n$ vary, this second experiment does the opposite by gradually reducing the time limit.
\end{itemize}

\subsubsection{Settings}

Experiment 1 used TSPLIB-CH130 to generate instances of up to 130 cities. We first thought it would be useful in Experiment 2 to have as many cities as we wished (which turned out to be useless) and, also, to investigate whether the experimental results change when another method is used to generate instances. For both these reasons, we have now changed the considered instances by assuming that the places to be visited are clients in a city -- even if we keep calling these places ``cities'' and using the Euclidean distance.
For this purpose, we first set the seed of the Pseudo Random Number Generator of AnyLogic (``{\sf getEngine().getDefaultRan\-domGene\-rator().set\-Se\-ed(main.\-instance);}'' where Parameter ``{\sf main.\-instance}'' is an integer in the ``{\sf main}'' of AnyLogic; The value of this parameter is the same for all mechanisms when they solve this particular in\-stance. Next, we generate the location of each city in polar coordinates (``{\sf double r=normal(150,0);}'' and ``{\sf double theta =uni\-form(0,360);}''), then translate them into AnyLogic's Cartesian coordinates (``{\sf setXY(main.rectan\-gleWidth/2+ r*cos(theta), main.\-rectangleHeight/2+r*sin(theta));}''). We generate 26 instances (instead of 130 in Experiment 2) of $n=121$ cities (instead of the quantity indicated on the horizontal axis in Figure \ref{fig-ratios-30min-centrb}, $i.e.,$ between $n=10$ and $n=55$ for $m=9$ salesmen).
Thus, every point in Figure~\ref{fig-varying-time} is the (fifth or ninth) decile of the total route length found by a given mechanism running 26 times (once per value of ``{\sf main.instance}'') divided by the total route length found by {\sf Centr$_b$} on these same 26 instances.
This figure shows the two successive parts in Experiment 2 together:
\begin{enumerate}
	\item \emph{Screening}: We start with a time limit of 30 minutes (1,800 seconds) and gradually divide it by two times ten times. Hence, the horizontal axis in Figure~\ref{fig-varying-time} is logarithmic.
	(Technically, in a ``Parameter Variation'' experiment, AnyLogic provides the integer ``{\sf index}'' to identify what iteration it is simulating. We set Parameter ``{\sf main.instance}'' in our model to ``{\sf (int)(index/10)}'' and Parameter ``{\sf main.maxi\-mumCom\-pu\-ta\-tionTimeSpan} to ``{\sf 1800000/Math.pow(2,in\-dex\%10)}''.)
	\item \emph{Detailed}: The points obtained in the screening between 3.5 seconds and 14 seconds are close to 1. This means that {\sf Centr$_b$} is the worst mechaninism when the time limit is below 3.5 seconds, and the best when this limit is greater than 14 seconds. Consequently, we conduct more exhaustive experiments between these two limits, starting at 14 seconds and progressively decreasing by 1.25 seconds ten times.
	(Technically, Parameter ``{\sf main.instance}'' is the same as in the previous paragraph, but ``{\sf maximumComputationTimeSpan} is set to ``{\sf 1800000/Math.pow(2,7) - 1250*(1+index\%10)}''.)
\end{enumerate}

The softwares and hardwares in Experiment 2 are the same as in Experiment 1, except that we updated CPLEX to version 20.10.

\subsubsection{Results}

\begin{figure}
	\begin{center}
	\includegraphics[width=8.5cm,height=10.7cm]{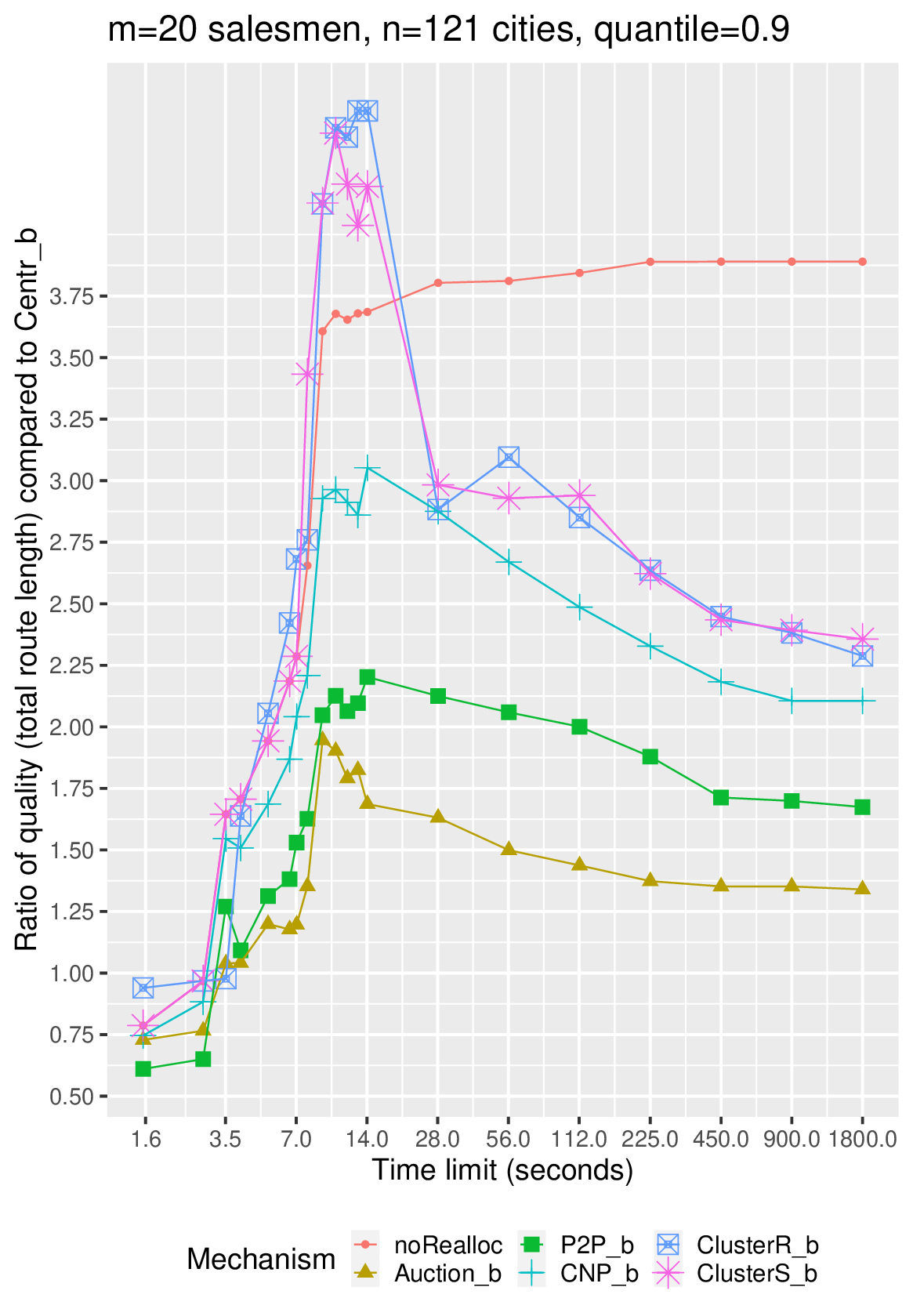}
	\includegraphics[width=8.5cm,height=10.7cm]{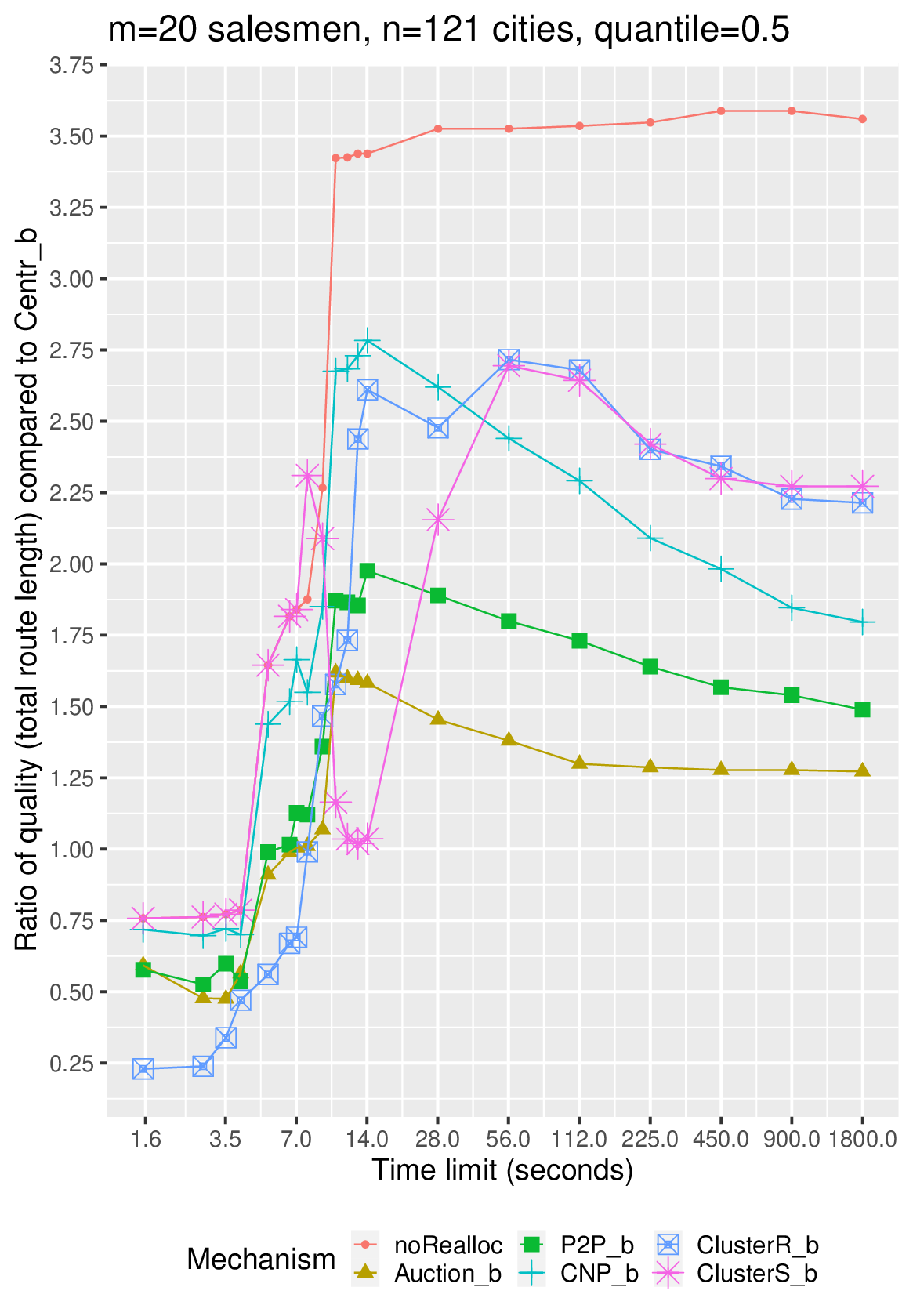}
	\caption{Experiment 2: Deciles of the ratios of quality of every mechanism compared to Mechanism {\sf Centr$_b$} for $m=9$ salesmen and $n=121$ cities generated from a normal distribution.}
	\label{fig-varying-time}
	\end{center}
\end{figure}

The first thing to notice in Figure~\ref{fig-varying-time} is that the ranking of the mechanisms for time limits between 28 seconds\footnote{Since the ninth deciles of {\sf ClusterR$_b$} and {\sf ClusterS$_b$} are sometimes larger than the deciles of {\sf NoRealloc}, we write 28 seconds instead of the above 14 seconds.} and 30 minutes is the same as in Experiment~1. 
Consequently, changing the location of the cities has not affected the ranking.
Therefore, the right-hand side of the first row in Table \ref{tab:comparison} is also true for Experiment~2.

The points for time limits below 7 seconds provide a ranking for the left-hand side entry of the first line in Table \ref{tab:comparison}, for which our previous article and Experiment 1 have no data. We only look at medians here because, for the first time in our experiments on both MTSP$_\textrm{b}$ and MTSP$_\textrm{s}$, the ninth deciles give a different ranking than medians.
It is also the first time that {\sf ClusterR$_s$} and {\sf ClusterS$_s$} incur different efficiencies. More precisely, {\sf ClusterR$_s$} is the best of all mechanisms, and {\sf ClusterS$_s$} is the worst decentralised mechanisms even if it is still better than {\sf CentrS$_s$}.

We find once \emph{again that our decentralised mechanisms are either all better or all worse than the centralised mechanism {\sf Centr$_\textrm{b}$}}, except in the tiny zone between 7 and 28 seconds where this is not so clear cut.


Understanding the difference of rankings between the selfish and benevolent versions of our mechanisms is important as it contradicts the intuition that centralisation causes (or, at least, is correlated with) efficiency.
We assume that this gap is due to the fact that two-city bundles are not taken into account by any of our selfish mechanisms.
From a more general point of view, \emph{despite a same level of (de-)centralisation and even an identical AnyLogic state chart, benevolent and selfish versions of a same mechanism may operate very differently}, as pointed out in Footnotes \ref{foot1} and \ref{foot2} which suggest variants of two of our mechanisms.

\section{Conclusion}
\label{sec:conclusion}

This article contributes to understanding when and how to (de-)centralise decision making in organisations. For this purpose, it compares mechanisms with various levels of (de-)cen\-tral\-i\-sa\-tion and that solves a Multiple Travelling Salesmen Problem (MTSP) in order to explore when and how to (de-)centralise allocation.
We study the efficiency of decisions when time is limited because the advantage of centralisation is said to be its capability to guarantee optimality in the absence of such a time limit, while the advantage of decentralisation would be its reactivity when this limit is tight.
We compare the efficiency of our mechanisms with one another, and also refer to the same comparison on the variant of MTSP in a previous article of ours in which the salesmen were assumed to be selfish. Our results indicate that the relative efficiency of the mechanisms is different in the traditional MTSP and the variant with selfish salesmen.
Our results with the traditional MTSP are therefore not as intuitive as what we obtained with this variant, in which a higher level of centralisation caused (or was just correlated with) a greater efficiency.
For both versions of MTSP, we think that an important conclusion is that our fully centralised mechanism -- which we call {\sf Centr} -- is either better than all our decentralised mechanisms, whatever their level of decentralisation, or worse than them all, and the zone in between is very narrow. In other words, {\sf Centr} is the best when the time frame is long enough to find the optimum or to approximate it, but suddenly becomes the worst when we gradually reduce the time available to make decisions.
Our results also highlight the fact that various mechanisms may be instances of a same decentralised organisation, and various mechanism designs for a same organisation may yield very different efficiencies.

This article addresses two versions of MTSP. In future work we would like to explore the impact of the level of (de-)cen\-tral\-i\-sa\-tion in a third variant of MTSP. We are interested in models for food delivery, to check if centralisation is more efficient in terms of kilometres travelled by delivery bikers, than a decentralised mobile application. If centralisation is best, then the difference of efficiency would provide insights into the value provided to bikers by companies currently operating such services.

\section{acknowledgements}
The author thanks the department of Industrial Engineering of INSA-Lyon for the use of the computers in one of its student laboratories to generate the large quantity of experimental data summarised in this article.

\section{Conflict of interest}
The author declares that he has no conflict of interest.

\bibliographystyle{spbasic}		
\bibliography{biblio,biblio2,biblio3}

\end{document}